\pdfoutput=1
\documentclass[a4paper,11pt]{article}
 \usepackage{lscape}
 \usepackage[all]{xy} 
 \usepackage{longtable} 
 \usepackage{jheppub} 
 \usepackage{graphicx}
 \usepackage{amssymb}
 \usepackage{amsmath}
\usepackage{mathtools}
\usepackage{listings}
\usepackage{dcolumn}
\usepackage{bm}
\usepackage{color}
\usepackage{multirow}
\usepackage{upquote}

\allowdisplaybreaks

\usepackage[table]{xcolor}

\definecolor{mag}{RGB}{255,0,255}
\definecolor{cy}{RGB}{0,255,255}
\definecolor{blu}{RGB}{51,51,255}

\newcommand{\lsim}{{\;\raise0.3ex\hbox{$<$\kern-0.75em\raise-1.1ex\hbox{$\sim$}}\;}}
\newcommand{\gsim}{{\;\raise0.3ex\hbox{$>$\kern-0.75em\raise-1.1ex\hbox{$\sim$}}\;}}
\newcommand{\beq}{\begin{equation}}
\newcommand{\eeq}{\end{equation}}
\newcommand{\bea}{\begin{eqnarray}}
\newcommand{\eea}{\end{eqnarray}}

\mathchardef\minus="002D

\title{\boldmath Addressing Astrophysical and Cosmological Problems With Secretly Asymmetric Dark Matter}
\author[a]{Christopher Dessert,}
\author[b]{Can Kilic,}
\author[b]{Cynthia Trendafilova,}
\author[c]{and Yuhsin Tsai}

\affiliation[a]{Leinweber Center for Theoretical Physics, Physics Department, University of Michigan, Ann Arbor, MI, 48109} 
\affiliation[b]{Theory Group, Department of Physics, The University of Texas at Austin, Austin, TX 78712} 
\affiliation[c]{Maryland Center for Fundamental Physics, University of Maryland, College Park MD 20742}

\abstract{We present a simple model of dark matter that can address astrophysical and cosmological puzzles across a wide range of scales. The model is an application of the Secretly Asymmetric Dark Matter mechanism, where several flavors of dark matter are fully asymmetric despite an exact dark matter number symmetry $U(1)_{\chi}$. The dark matter relic abundance arises from these asymmetries, generated in the early universe through right-handed neutrino decays. The $U(1)_{\chi}$ is gauged by a massless dark photon, and asymmetries with opposite signs in the different DM flavors result in the formation of bound states. Dark acoustic oscillations in the early universe lead to a suppression in the matter power spectrum for addressing the $\sigma_8$ problem. The dark photon as a relativistic degree of freedom contributes to $\Delta N_{eff}$, easing the tension between the CMB and low redshift $H_{0}$ measurements. Finally, scattering between the bound states after structure formation leads to a flattening of the dark matter distribution at the cores of haloes.}

\preprint{
\begin{flushleft} 
UTTG-23-18\\
LCTP-18-24\\
\end{flushleft} 
}

\begin{document} 
\maketitle
\flushbottom

\section{Introduction}
\label{sec:introduction}

Secretly Asymmetric Dark Matter (SADM) has recently been proposed~\cite{Agrawal:2016uwf} as a mechanism to generate the dark matter (DM) relic abundance through an asymmetry in the dark sector despite an exact (global or gauged) DM number symmetry $U(1)_{\chi}$. In the implementation of this idea in ref.~\cite{Agrawal:2016uwf}, the asymmetry is first generated in the visible sector through high-scale leptogenesis~\cite{Fukugita:1986hr}\footnote{See reviews of leptogenesis such as ref.s~\cite{Davidson:2008bu,Fong:2013wr} for a comprehensive list of references.} and then transferred to a dark sector containing three DM flavors $\chi_{i}$, a flavorless mediator $\phi$, and a dark photon $\gamma_{D}$ via a coupling to the right-handed leptons of the Standard Model (SM). The mediator in that case is charged both under SM hypercharge and under $U(1)_{\chi}$, and therefore leads to mixing between the photon and the dark photon at the one-loop level. This is phenomenologically only acceptable if the dark photon has a nonzero mass~\cite{Chang:2018rso}, and therefore the experimental signatures of the model manifest themselves at short distances, but in the context of long distance physics it falls under the collisionless cold DM category.

In this work we introduce a different implementation of the SADM mechanism, where a massless dark photon is phenomenologically allowed, and leads to a rich set of astrophysical signatures, despite the rather minimal high-energy setup. We argue that this setup is capable of addressing open questions in astrophysics both at the large (Hubble-scale) and small (galaxy-scale) scales simultaneously. The model we propose serves as a short-distance implementation of two mechanisms that have recently been proposed to address the same open questions, namely Dark Acoustic Oscillations~\cite{CyrRacine:2012fz,Chacko:2016kgg,Raveri:2017jto,Buen-Abad:2017gxg,Pan:2018zha} and Hidden Hydrogen DM~\cite{Boddy:2016bbu,Chacko:2018vss}. By focusing on a particular region of parameter space of the model, we show that the $H_{0}$ discrepancy~\cite{Bernal:2016gxb}, the $\sigma_{8}$  discrepancy~\cite{2017MNRAS.465.1454H,Kohlinger:2017sxk,Joudaki:2017zdt,Troxel:2017xyo,Persic:1995ru} and the mass deficit problem~\cite{Moore:1994yx,Persic:1995ru,2013ApJ...765...25N} in dwarf galaxies and galaxy clusters can all be addressed. There have been many efforts in the literature to address these discrepancies using non-standard DM self interactions~\cite{Cline:2012is,Foot:2014uba,Chacko:2015noa,Buen-Abad:2015ova,Lesgourgues:2015wza, Ko:2016uft,Ko:2016fcd,Prilepina:2016rlq,Agrawal:2017rvu,Baldes:2017gzu}. The SADM model presented here extends these efforts by providing a minimal short-distance setup that nevertheless has rich enough dynamics to address multiple issues across a wide range of scales.

The core feature of the SADM mechanism is that due to the unbroken DM number symmetry, the total dark charge of the universe is zero at all times, but there can be several DM flavors that have individual asymmetries. Due to the crucial role of there being multiple DM flavors, SADM can be considered as a special case of the Flavored Dark Matter scenario~\cite{MarchRussell:2009aq,Batell:2011tc,Agrawal:2011ze,Kile:2013ola,Agrawal:2014aoa,Chen:2015jkt,Agrawal:2015kje}. Unlike the model studied in ref.~\cite{Agrawal:2016uwf} however, where the heavier DM flavors decay to the lightest one before Big Bang Nucleosynthesis (BBN) and result in a symmetric distribution of the lightest DM flavor at late times, in the model we study in this work all DM flavors are extremely long-lived and are still present today. This means that flavors with opposite signs of asymmetry (and with opposite dark charges) must coexist, along with a massless dark photon that mediates long-range interactions, and therefore multiple bound states can form, each one behaving as atomic DM~\cite{1986PhLB..174..151G,Kaplan:2009de,Cline:2012is}. These bound states, along with unbound DM particles and the massless dark photon then give rise to rich dynamics across a range of distance scales. The dark photon as an additional relativistic degree of freedom helps ease the tension between CMB-based~\cite{Aghanim:2018eyx} and low redshift~\cite{Bernal:2016gxb,Bonvin:2016crt,Riess:2018byc} measurements of $H_{0}$, dark acoustic oscillations in the early universe that arise from the scattering of free and bound states leads to a suppression of the matter power spectrum at small scales, and the scattering between bound states after structure formation leads to a flattening of the DM density distribution inside haloes.

The layout of this paper is as follows: In section~\ref{sec:model} we introduce the short-distance model, and we consider its phenomenology in the context of particle physics observables and constraints. In section~\ref{sec:cosmo} we then study the cosmological features of the model, and the astrophysical signatures that it gives rise to at the large and small scales. We conclude in section~\ref{sec:conclusions} and we consider directions for future research.

\section{The Model, and particle physics considerations}
\label{sec:model}

In this section we will introduce our SADM model and consider its high energy aspects. We assume heavy right handed neutrinos $N_{i}$ exist and that they have Majorana masses as well as a Yukawa coupling to the SM leptons
\beq
{\mathcal L}\supset M_{N,ij}N_{i}N_{j}+Y^{N}_{ij}N_{i}\ell_{j}H^{\dag}+{\rm h.c.},
\eeq
where $H$ is the SM Higgs doublet. This well-studied extension of the SM generates both light neutrino masses through the see-saw mechanism and it also allows the creation of a net $B-L$ number as the origin of the matter-antimatter asymmetry in the SM sector through CP violating phases in the coupling matrix $Y^{N}_{ij}$. These features are explored in depth in reviews of leptogenesis~\cite{Davidson:2008bu,Fong:2013wr}.

To this setup we add three flavors of Dirac fermions $\left(\chi,\chi^{c}\right)_{i}$ and a scalar $\phi$ with the couplings (in the $\chi$-mass basis)
\beq
{\mathcal L}\supset M^{2}_{\phi}|\phi|^{2}+m_{\chi,i}\chi^{c}_{i}\chi_{i}+\lambda_{ij}N_{i}\chi_{j}\phi+{\rm h.c.}.
\label{eq:FDM}
\eeq
Note that an exact $U(1)_{\chi}$ symmetry exists, under which all $\chi$ have charge +1, while all $\chi^{c}$ and $\phi$ have charge -1. We will take this symmetry to be gauged by the dark photon, with a fine structure constant $\alpha_{d}$. The $\lambda_{ij}$ coupling matrix contains physical phases and is a source of CP-violation in the dark sector. We choose the $N$ to only couple to the left-handed $\chi$ (and $\phi$) but not the right-handed $\chi^{c}$ (and $\phi^{*}$). If couplings to both $\chi$ and $\chi^{c}$ are present with ${\mathcal O}(1)$ coefficients, this can allow a heavier $\chi$ particle to decay to a lighter one within the age of the universe, as will be discussed later in this section. While a coupling to $\chi^{c}$ will be generated through quantum effects, this effect is suppressed by ratio of $\chi$ and $N$ masses, and this does not invalidate our estimate of the $\chi$ lifetime presented below. We will soon introduce a convention for assigning the flavor labels $i$. 

The mediator $\phi$, being a scalar, will be taken to be heavy, though we will assume that it is lighter than the right-handed neutrinos. Once the right-handed neutrinos become non-relativistic and the interaction of equation~\ref{eq:FDM} drops out of equilibrium, the SM and dark sectors decouple from one another. As the heavy neutrinos $N$ decay, they will then generate asymmetries both in the SM leptons and in the dark sector (in $\phi$ and the $\chi_{i}$); however as mentioned before $U(1)_{\chi}$ is never broken and therefore the total dark charge of the universe is zero at all times. With the interaction of equation~\ref{eq:FDM} out of equilibrium, the asymmetries generated in the three $\chi$ flavors cannot wash each other out. As the temperature drops further, the $\phi$ particles become nonrelativistic and the symmetric part of the $\phi$ particle distribution annihilates to dark photons. The asymmetric $\phi$ particles then decay as $\phi\rightarrow \chi\, \ell^{\pm} H^{\mp}$, transferring their dark charge to the $\chi$ flavors, such that after the $\phi$ decays the asymmetry in the dark sector will reside entirely in the three DM flavors, in such a way that the total dark charge of the universe remains zero.  Note that the heaviness of both $\phi$ and $N$ means that the $\chi$ are very challenging to access in collider, direct detection, or indirect detection experiments. Our model does however have astrophysical signatures, which are the focus of this paper and which we will study in the next section.

Assuming generic phases in the entries of the $Y^{N}_{ij}$ and the $\lambda_{ij}$ matrices, the ratio of the typical magnitudes of the elements of $Y^{N}_{ij}$ and the typical magnitudes of the elements of $\lambda_{ij}$ will determine the branching ratio of the decaying $N$ into the visible vs the dark sector, and thereby the ratio of the overall lepton number (more precisely, $B-L$ number) and the flavor-by-flavor $\chi$ asymmetries that are generated through heavy neutrino decays.

Therefore, for similar magnitudes in the entries of the $Y^{N}_{ij}$ and the $\lambda_{ij}$ matrices, the fact that $\Omega_{DM}\sim 5\Omega_{B}$ suggests the GeV scale for the mass of the heaviest $\chi$ flavor, but if the entries of one coupling matrix are larger than those of the other, then this mass can also vary in either direction. As we will show in section~\ref{sec:cosmo}, the region of parameter space that is of interest for us has the mass of the heaviest DM flavor $\sim{\mathcal O}(10)$~GeV, the mass of the lightest DM flavor $\sim{\mathcal O}(1)$~MeV, and a value for $\alpha_{d}$ of order several percent. In this parameter region, the symmetric parts of the $\chi$ distributions annihilate efficiently, leaving behind only the asymmetric part generated during the $N$ decays.

\begin{figure}
\begin{center}
\includegraphics[width=0.6\textwidth]{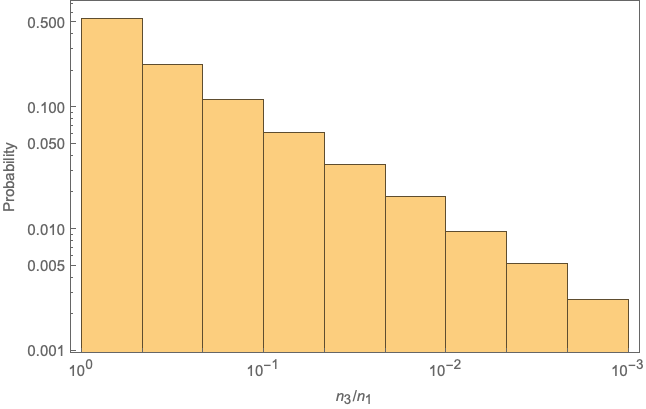}
\end{center}
\caption{The estimate of the probability distribution function for the ratio $n_{3}/n_{1}$. See main text for the details of the calculation.}
\label{fig:xrpdf}
\end{figure}

The asymmetry generated in the different $\chi$ flavors will differ from one another. In fact, due to the total dark charge of the universe being zero, there will always be one $\chi$ flavor with one sign of the asymmetry, and two flavors with the opposite sign, after the $\phi$ particles have all decayed. We will adopt the following naming convention for the three $\chi$ flavors: The $\chi$ flavor which is unique in its sign of the asymmetry will be labeled $\chi_{1}$, whereas the two $\chi$ flavors with the other sign of the asymmetry will be labeled $\chi_{2}$ and $\chi_{3}$, such that $m_{\chi_{2}}>m_{\chi_{3}}$. As we will see in the next section, the most interesting phenomenology is obtained when $\chi_{3}$ and not $\chi_{1}$ is the lightest DM flavor. Thus we will take $m_{\chi_{1}}>m_{\chi_{3}}$ as well, while either one of $\chi_{1}$ or $\chi_{2}$ could be the heaviest flavor. The possible bound states are then formed between $\chi_{1}$ and $\chi_{2}$, labeled $H_{12}$, and between $\chi_{1}$ and $\chi_{3}$, labeled $H_{13}$. Since the total dark charge of the universe is zero at all times, the asymmetries of $\chi_{2}$ and $\chi_{3}$ add up in magnitude to that of $\chi_{1}$.

The overall size of the entries of the $\lambda_{ij}$ coupling matrix determines the branching fraction of $N$ to the dark sector, and therefore the overall size of the asymmetries of all $\chi$ flavors, while the specific flavor and phase structure in the coupling matrix determines the branching fractions among the DM flavors. The three comoving asymmetries $\Delta Y_{i}$ of the $\chi$ flavors (after $\phi$ decays) satisfy $\sum_{i=1}^{3}\Delta Y_{i}=0$. In terms of the physical number densities of the three $\chi$ particles (or antiparticles, depending on the sign of the asymmetry), this can also be written as $n_{1}=n_{2}+n_{3}$.  Thus the $n_{i}$ are not independent, however the ratio $n_{3}/n_{1}$ can in principle have any value in the interval $[0,1]$.

We can estimate the probability distribution for this ratio using Monte Carlo methods with a prior where each entry of the $3\times3$ matrix $\lambda_{ij}$ is randomly chosen over the unit complex circle. To calculate the asymmetry generated in each flavor from a given matrix $\lambda_{ij}$, we follow standard techniques from leptogenesis~\cite{Davidson:2008bu}. We assume the heavy neutrino masses are hierarchical with $m_{N_{1}} \ll m_{N_{2}}, m_{N_{3}}$ (we use $m_{N_{1}} = 1.0\times10^{16}$~GeV as a benchmark value). $N_{1}$ decays are fast compared to the Hubble scale, and therefore the asymmetry production takes place in the strong washout regime. Ratios of the $\Delta Y_{i}$ generated in the $N_{1}$ decays can be obtained from the ratios of the CP asymmetry factors for each flavor $i$ 
\beq
\epsilon_{ii} \propto \mathrm{Im}\lbrace[\lambda]_{1i}[m^* \lambda]_{1i}\rbrace,
\eeq
where the matrix $[m]_{ij}$ is given by
\beq
[m]_{ij} = [\lambda]_{ki} [\lambda]_{kj} / m_{N_{k}}.
\eeq
For the branching ratios of the subsequent $\phi$ decays, only the tree level contributions with an off-shell $N_{1}$ are used, since any additional CP violation introduced at this stage is subdominant. Note also that an {\it overall} rescaling of all $\lambda_{ij}$ entries does not affect the probability distribution of $n_{3}/n_{1}$, and therefore we are not committing ourselves to a particular size of the $\lambda_{ij}$ couplings by generating random numbers over the complex unit circle. For a given choice of $m_{N_{1}}$ and $m_{\chi_i}$, such a rescaling can be chosen such that the correct overall $\rho_{DM}$ and $\rho_{B}$ are obtained. The result for the $n_{3}/n_{1}$ probability distribution is shown in figure~\ref{fig:xrpdf}. As we will see in the next section, the region of greatest phenomenological interest corresponds to $n_{3} / n_{1} \lsim 0.1$, which does not require significant fine-tuning.

\begin{figure}
\begin{center}
\includegraphics[width=0.4\textwidth]{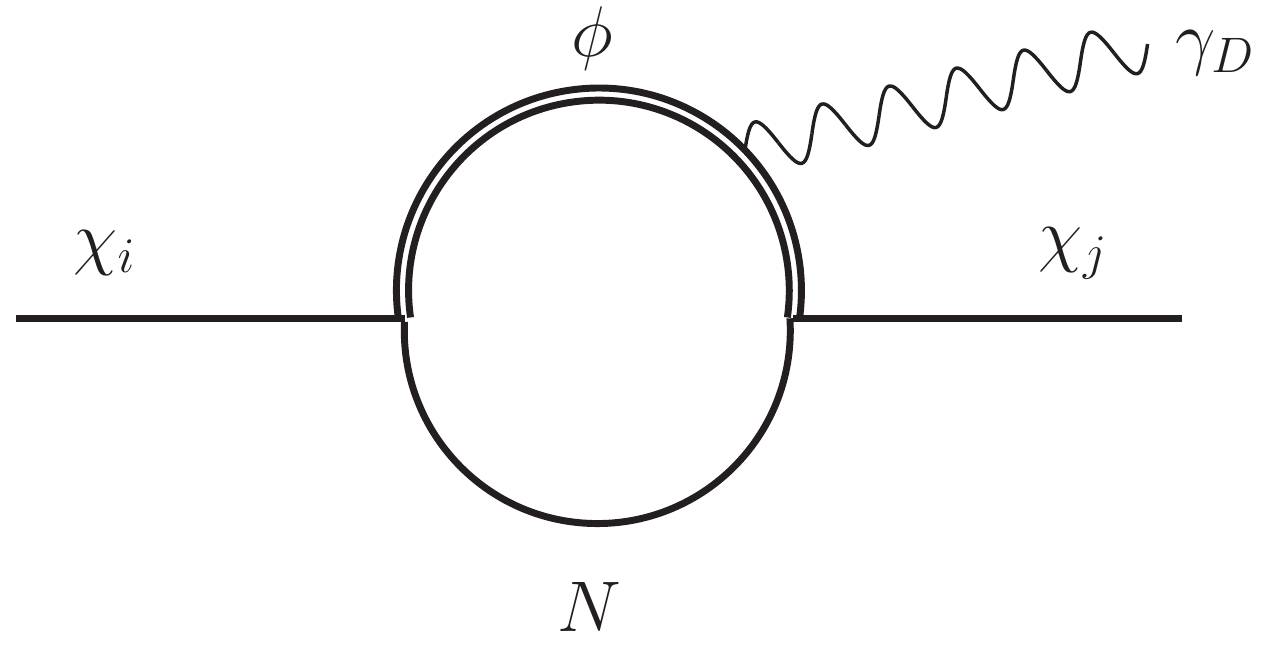}
\end{center}
\caption{Decay mode for a heavier DM flavor to decay to a lighter one.}
\label{fig:decay}
\end{figure}

Even without a compressed spectrum among the DM flavors, all $\chi$ are extremely long-lived, and are still undecayed today. The leading decay mode arises from the one-loop diagram shown in figure~\ref{fig:decay}. Let us estimate the lifetime associated with this decay mode. Since $N$ is the heaviest particle in the spectrum, we integrate it out first. At dimension 5, this generates the operator (we use $\lambda$ to stand in for any $\mathcal{O}(1)$ entries in the $\lambda_{ij}$ coupling matrix, and we are suppressing flavor indices on the $\chi$)
\beq
{\mathcal O}_{5}=\frac{\lambda^{2}}{m_{N}}\left(\chi\chi\phi^{2}+{\rm h.c.}\right).\label{eq:o5}
\eeq
Note however that this operator does not contribute to the decay mode of figure~\ref{fig:decay}, since one needs a factor of $\phi\phi^{*}$ in order to close the $\phi$ loop. In order to do this, we need to go to the dimension 6 operator generated from integrating out $N$
\beq
{\mathcal O}_{6}=\frac{\lambda^{2}}{m_{N}^{2}}\left(\bar{\chi}\bar{\sigma}^{\mu}\partial_{\mu}\chi\phi\phi^{*}\right).
\eeq
Next, we will integrate out the heavy mediator $\phi$. In the low energy theory below $m_{\phi}$, the effective operator that leads to the $\chi$ decay is the operator
\beq
\bar{\chi}\bar{\sigma}^{\mu}\partial^{\nu}\chi\, F_{d\,\mu\nu},
\eeq
where $F_{d\,\mu\nu}$ is the field strength for $U(1)_{\chi}$. Since all derivatives in this effective operator act on the external momenta, which are of order $m_{\chi}$, the decay rate can be estimated as
\beq
\Gamma_{\rm approx.}=\frac{1}{8\pi}\left(\frac{\lambda^{2}e_{d}}{16\pi^{2}m_{N}^{2}}\right)^{2}m_{\chi}^{5}\\
\eeq
We have also performed the full loop calculation for this process, and we find complete agreement with the effective field theory estimate presented above, with an added numerical factor of $1/72$. The full answer for the decay rate is thus found to be, up to corrections of order $\left(m_{\phi}^{2} / m_{N}^{2}\right)$ and of order $\left(m_{\chi, {\rm final}}^{2} / m_{\chi, {\rm initial}}^{2}\right)$:
\begin{eqnarray}
\Gamma&=&\frac{1}{72}\Gamma_{\rm approx.}\label{eq:decaysub1}\\
 &\approx&(10^{-42}{\rm eV})\left(\lambda^{4}e_{d}^{2}\right)\left(\frac{m_{\chi}}{10 {\rm GeV}}\right)^{5}\left(\frac{10^{12}{\rm GeV}}{m_{N}}\right)^{4}\label{eq:decaysub2}\\
 &\ll&H_{0}\approx 10^{-33}~{\rm eV}.
 \label{eq:decay}
\end{eqnarray}
The lifetime is several orders of magnitude larger than the age of the universe for the parameter region that will be used in this paper. The leading tree-level decay diagrams are even further suppressed than this decay mode, and are therefore irrelevant. As discussed earlier in this section, if the UV theory contains couplings of $N$ to both $\chi$ and $\chi^{c}$ with ${\mathcal O}(1)$ coefficients, then the operator ${\mathcal O}_{5}$ of equation~\ref{eq:o5} will contain a cross-term containing $\chi\chi^{c}\phi\phi^{*}$ which can contribute to the decay process at order $1/m_{N}$ instead of at order $m_{\chi} / m_{N}^{2}$, leading to an unacceptably short lifetime. If the coupling to $\chi^{c}$ is absent in the UV theory however, it is only generated with an $m_{\chi} / m_{N}$ suppression, and this contributes at the same order as the piece already considered in equation~\ref{eq:decaysub1}. 

\begin{figure}
\begin{center}
\includegraphics[width=0.4\textwidth]{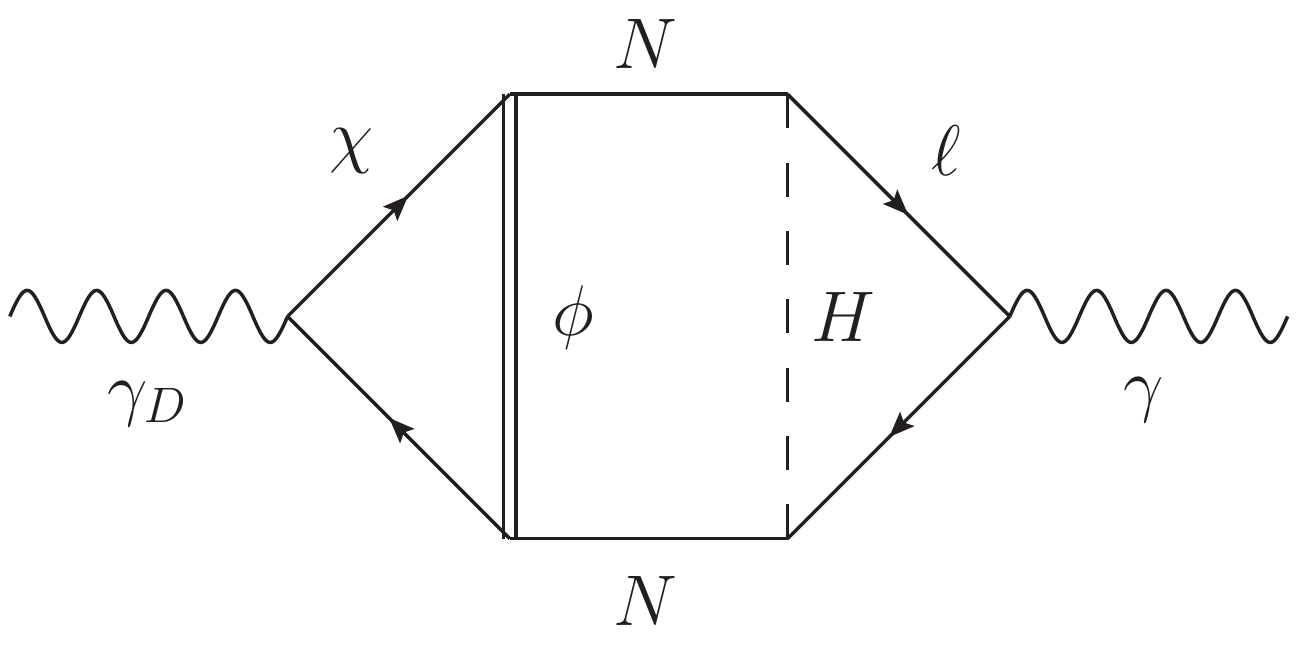}
\end{center}
\caption{Leading contribution to mixing between the dark photon and the SM photon.}
\label{fig:mixing}
\end{figure}

As we will see in the next section, the most interesting phenomenological features of the model will be obtained when the mass of the heaviest flavor ($\chi_{1}$ or $\chi_{2}$) is ${\mathcal O}(10)$~GeV, the mass of the next heaviest flavor is ${\mathcal O}(0.1-1)$~GeV, and the mass of the lightest flavor ($\chi_{3}$) is ${\mathcal O}(1)$~MeV. This means that $H_{12}$ will be strongly bound and difficult to ionize, serving as a neutral DM component, while $H_{13}$ may be easy to ionize, such that a non-negligible fraction of $\chi_{3}$ (and the corresponding amount of $\chi_{1}$) can be unbound at certain times during the early universe, delaying the decoupling of the dominant DM component from the dark radiation, and leading to interesting dynamical effects.

For a massless dark photon, it needs to be checked that the mixing with the SM photon is acceptably small. More precisely, in the basis where $\gamma_{D}$ only interacts with $\chi$ but not the SM fermions, the millicharge acquired by $\chi$ under electromagnetism needs to be small. The $\gamma$-$\gamma_{D}$ mixing is induced at the three-loop level: see figure~\ref{fig:mixing}. Matching on to an effective operator $\epsilon F^{\mu\nu}_{D}F_{\mu\nu}$, one can estimate
\beq
\epsilon\sim \frac{e e_{d} \lambda^{2} Y^{2}}{(16\pi^{2})^{3}}.
\eeq
If $\lambda^{2}\sim Y^{2}\sim 0.1$ and $\alpha_{d}\sim\alpha$, then $\epsilon$ is below $10^{-10}$. Let us compare this to the constraints for millicharged DM (figure 1 in ref.~\cite{Kadota:2016tqq}, with $m\sim{\mathcal O}(10)$~GeV, the mass of the heaviest DM flavor), keeping in mind that some of the constraints will be relaxed, as these were derived with the assumption that all DM particles are millicharged while we will focus in this work exclusively on the possibility that all but a small fraction of the $\chi$ particles are in bound states, and therefore have no {\it net} millicharge. In particular, the leading constraints from galactic and cluster magnetic fields probe the DM at small momenta, or at distance scales larger than the size of the bound states, thus these constraints cannot resolve the millicharges of the constituents. While the momentum scale that is probed at LUX is still too small to resolve the constituents of $H_{12}$, it may be sufficiently large to resolve the constituents of $H_{13}$. However, in the parameter space of interest to us, $H_{13}$ is the subdominant DM component, which relaxes the constraints below the generic values of $\epsilon$ in our model. Furthermore, for a significant fraction of the parameter region of interest to us (especially when $m_{\chi_{2}}>m_{\chi_{1}}$), the mass of $H_{13}$ is below the sensitivity of LUX. It is however interesting to note that future direct detection experiments with a low mass threshold may be able to test the scenario presented here. Finally, for the parameter region of interest, the value of $\epsilon$ in our model is also below the bounds from supernovae~\cite{Chang:2018rso}.

As a final consideration about the particle physics nature of our model, we want to address the potential concern that flavor oscillations between the DM flavors may cause a washout of the asymmetries, as the particles of $\chi_{1}$ oscillate into the antiparticles of $\chi_{2}$ or $\chi_{3}$, or the other way around. However, the large mass gap between the $\chi$ flavors in the parameter region of interest in this paper, and the extremely small intrinsic widths of these states makes oscillations so small that they are negligible for all intents and purposes.

\section{Cosmology and astrophysical signatures}
\label{sec:cosmo}

Having discussed the short-distance physics aspects of our model, we now turn our attention to the role it plays in the dynamics of the early universe, as well as in late-time astrophysical processes. Although the dark sector is hidden from direct experimental probes via non-gravitational interactions, the dynamics within the SADM sector can significantly change the gravitational potential in the early universe and it leaves visible signals on both the large and small scale structures. We will show that there is a region of parameter space where observational anomalies on scales of dwarf galaxies ($\sim$ kpc size)~\cite{Moore:1994yx,Persic:1995ru}, galaxy clusters ($\sim$ Mpc size)~\cite{2013ApJ...765...25N}, the $\sigma_8$ problem ($\sim10$ Mpc size)~\cite{2017MNRAS.465.1454H,Kohlinger:2017sxk,Joudaki:2017zdt,Troxel:2017xyo,Persic:1995ru}, and the tension in the measurements of $H_{0}$ from the CMB~\cite{Aghanim:2018eyx} and from low redshift measurements~\cite{Riess:2016jrr} may all be addressed. 

The input parameters of the model that are relevant at large distances can be listed as the comoving asymmetries of the $\chi$ flavors (of which only two are independent, and which can therefore be parameterized in terms of $n_{1}$ and $n_{3} / n_{1}$ once the symmetric part of the $\chi$ distributions annihilates away), the masses of the three $\chi$ flavors, and $\alpha_{d}$. In order to calculate how the temperature of the dark sector is related to that of the SM sector, consider energies above $m_{N}$ where the interaction of equation~\ref{eq:FDM} keeps the two sectors in equilibrium. As the temperature drops below $m_{N}$, the visible and dark sectors decouple. At this time, the dark sector contains the three Dirac fermions $\chi_{1,2,3}$, the complex scalar $\phi$, and the dark photon $\gamma_D$. $\phi$ is only somewhat lighter than the $N$, and for the parameter region of interest for us, the mass of the lightest flavor is ${\mathcal O}(1)$~MeV. Therefore, at the time of structure formation, only $\gamma_{D}$ is still relativistic. 

If there are no other degrees of freedom in the visible sector other than those of the SM up to energies of $m_{N}$, then by using the ratio of the number of relativistic degrees of freedom in the two sectors at the decoupling scale and at the scale of structure formation, one obtains $\Delta N_{eff}=0.75$. However, since the two sectors decouple at a very high temperature scale, one needs to take into account the possibility that the visible sector contains additional degrees of freedom (e.g. connected to the solution of the electroweak naturalness problem, grand unification, etc.) that release additional entropy in the visible sector as the universe cools down, and thereby further suppress $\Delta N_{eff}$. The value 0.75 should therefore be considered as an upper limit, with the actual value depending on other possible extensions of the SM. Since a value of 0.75 is outside the $2\sigma$ contour for reconciling the $H_{0}$ discrepancy (see e.g.,~\cite{Aghanim:2018eyx}), we will adopt a benchmark value of $\Delta N_{eff}=0.60$ for the remainder of the paper\footnote{According to Fig.~35 in~\cite{Aghanim:2018eyx}, this rather large $\Delta N_{eff}$ is within $2\sigma$ constraint from the joint Planck TT,TE,EE$+$lowE$+$lensing$+$BAO fit including the low redshift measurement~\cite{2018ApJ...855..136R}.}, which can be obtained for example if there are six Dirac fermions beyond the SM in the visible sector. Adding even more degrees of freedom to the visible sector would further reduce $\Delta N_{eff}$. It should also be noted that at the time of BBN, $\Delta N_{eff}$ is even smaller ($\Delta N_{eff}=0.42$) since $\chi_{3}$ has not yet become nonrelativistic at that time for $m_{\chi_{3}}\lsim$~MeV, which is the preferred value for addressing the $\sigma_{8}$ problem as we will see later. Thus our model is compatible with BBN constraints~\cite{Cooke:2013cba}. 

There are two recombination processes during the early universe, of $\chi_{1}$ with $\chi_{2}$ into $H_{12}$, and of the remaining $\chi_{1}$ with $\chi_{3}$ into $H_{13}$. $H_{12}$ forms earlier due to its larger binding energy, and the remaining $\chi_{1}$ and $\chi_{3}$ particles scatter with each other and remain in thermal equilibrium at this time. Similar to the proton-hydrogen scattering in the SM, the scattering cross section between the non-relativistic $\chi_{1}$ and the $H_{12}$ bound state is sufficient for keeping them in thermal equilibrium with each other. In the scenario we are interested in, the momentum transfer $q$ necessary to keep $H_{12}$ in thermal equilibrium at a dark temperature $T_d\sim 10$ eV during the dark acoustic oscillations is given by $q\sim \sqrt{m_{\chi_1}\,T_d}\sim 10^{4.5}$ eV, which is much smaller than the inverse size $a_0^{-1}\sim (\alpha_d\,m_{\chi_2})\sim 10^{7}$ eV of the $H_{12}$ bound state (for $m_{\chi_2}\sim$ GeV). In this limit, the scattering rate of $H_{12}$ and $\chi_{1}$ can be estimated as $\sigma v_{\chi_1}n_{\chi_1}\sim 4\pi\alpha_d^2m_{\chi_2}^2a_0^{4}\sqrt{T_d/m_{\chi_1}}n_{\chi_1}\sim 10^{-26}$ eV~\cite{Cline:2012is}, which exceeds the Hubble expansion rate at that time. We therefore treat $H_{12}$-$\chi_{1}$ to be in thermal equilibrium during the acoustic oscillations.
While $\chi_{1}$-$\chi_{3}$ scattering is also efficient, the entire dark sector is then in thermal equilibrium with the dark photon, and the resulting DM oscillations delay the formation of large scale structure.

The oscillation stops as $H_{13}$ recombines, which leaves too few free (dark-)charged particles to sustain the oscillations. These dark acoustic oscillations generate a small but visible damping of the matter power spectrum and may provide a solution to the $\sigma_8$ problem. A more complete parameter fitting procedure including also the CMB and BAO data is necessary to confirm this claim in full detail, however in this work we will take a simpler approach and we will calculate the size of power spectrum suppression to argue that the claim is plausible, leaving a more detailed analysis to future work.

We will also show that at later times, during the formation of DM halos, the scattering between $H_{12}$ bound states leads to thermal equilibrium and provides a solution to the core/cusp problem of dwarf galaxies. The non-trivial velocity dependence due to the inelastic scattering $H_{12}H_{12}\to H_{12}H_{12}\,\gamma_{D}$ through the dark hyperfine transition also gives the right cross section for cores to form in relaxed cluster halos. As we will see, achieving this will favor the parameter region where the heaviest DM flavor mass is $\sim{\mathcal O}(10)$~GeV, where the value of $\alpha_{d}$ is a few percent, and $n_{3} / n_{1}\sim 0.1$. Furthermore, we will see that achieving the correct amount of damping of large scale structure favors the MeV range for $m_{\chi_{3}}$. Thus there is a region of the parameter space of our model where all parameters are technically natural and where all three structure formation problems (dwarf, cluster and $\sigma_{8}$) and the $H_{0}$ tension could be addressed. Below, we explore each of these aspects of our model in detail.

\subsection{Dark recombination(s)}\label{sec:rec}
The formation of the $H_{12}$ and $H_{13}$ bound states plays an important role in structure formation. In particular dark acoustic oscillations end when there are no longer sufficiently many unbound $\chi$ particles left to sustain them. Similar to the recombination of hydrogen in the SM, for both $H_{12}$ and $H_{13}$, recombination proceeds through the formation of the excited states $(n\geq 2)$, with a subsequent decay into the ground state either through a two photon emission or through a Lyman-$\alpha$ transition, where the dark photon becomes redshifted before ionizing other bound states. 

A brief note on notation: we have introduced $n_{i}$ in the previous section to denote the physical densities in each flavor. Since in this section we will also need to keep track of just the density of unbound particles in each flavor, we will introduce the notation $n^{f}_{i}$, where $f$ stands for ``free''. With this definition $n_{1}=n^{f}_{1}+n_{H_{12}}+n_{H_{13}}$, $n_{2}=n^{f}_{2}+n_{H_{12}}$ and $n_{3}=n^{f}_{3}+n_{H_{13}}$. We also define the ionization fraction in each flavor as $X_{i}\equiv n^{f}_{i}/n_{i}$. Note that the dark charge neutrality of the universe enforces both $n_{1}=n_{2}+n_{3}$ and also $n^{f}_{1}=n^{f}_{2}+n^{f}_{3}$. 

\begin{figure}
\begin{center}
\includegraphics[width=7.6cm]{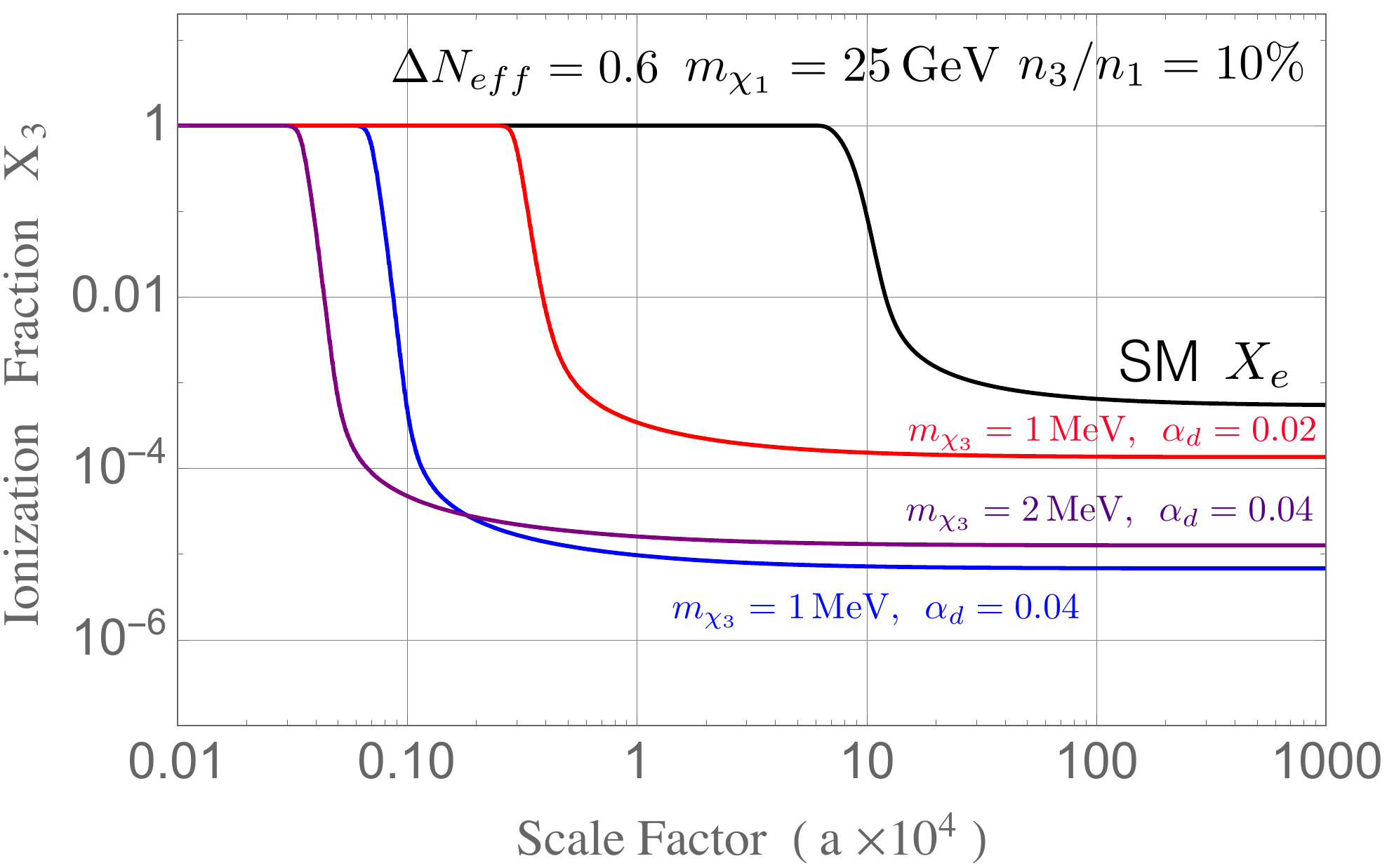}\,\,\,\includegraphics[width=7.6cm]{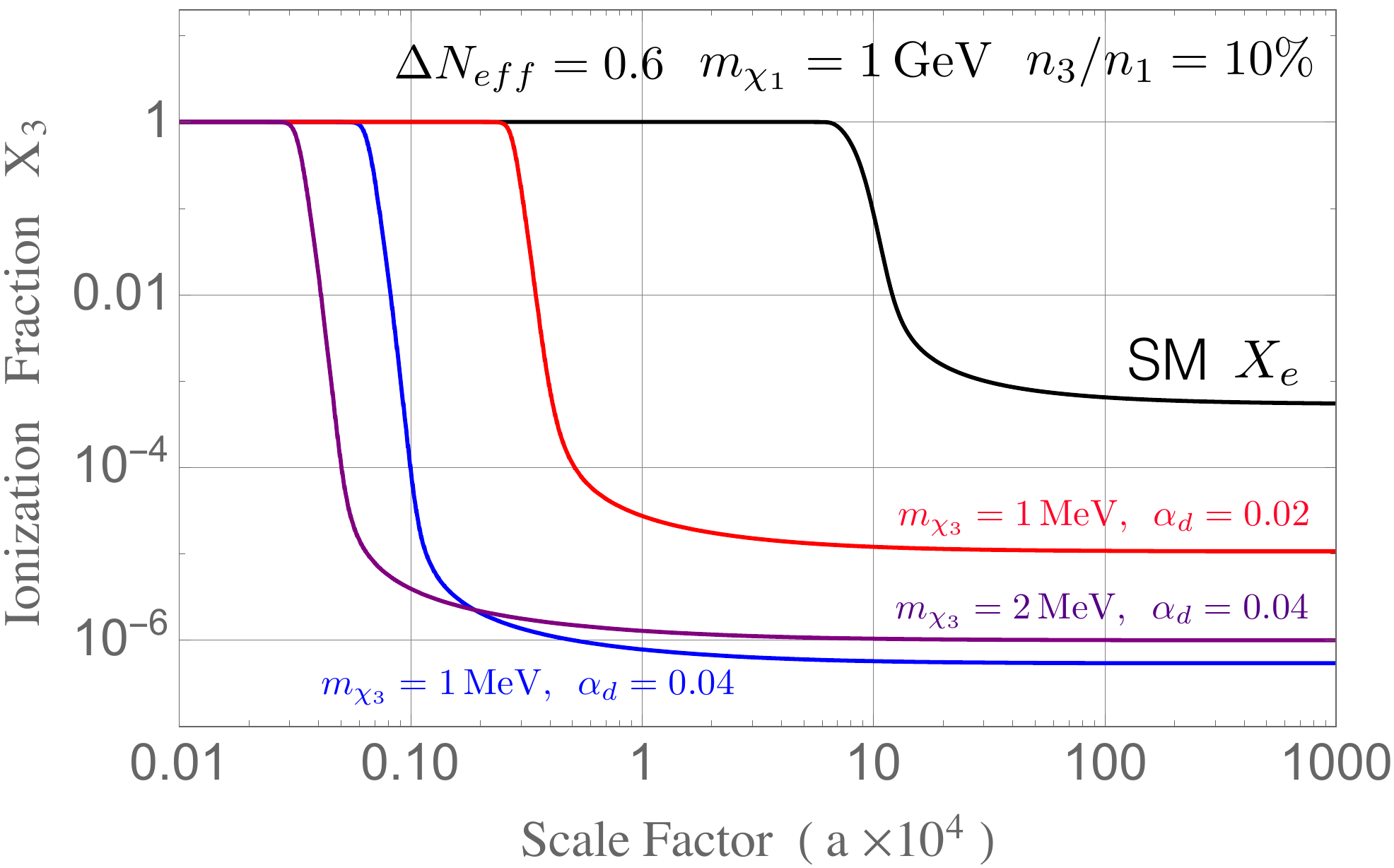}
\caption{The SM (black) and $\chi_{3}$ ionization fractions during the recombination epoch as a function of the scale factor, for representative values of $(m_{\chi_{3}},\,\alpha_{d})$ and two $\chi_1$ masses corresponding to the cases where $\chi_{1}$ is the heaviest DM flavor (left), and the second heaviest DM flavor (right).}\label{fig:recomb}
\end{center}
\end{figure}

The Boltzmann equation for $X_{2,3}$, can be written as~\cite{Chacko:2018vss}
\begin{equation}\label{eq:ionization}
\frac{dX_{2,3}}{dt}=C_{2,3}\Big\{(1-X_{2,3})\beta_{2,3}-X_{2,3}\,n^{f}_{1}\alpha^{(2)}_{2,3}\Big\},\quad\beta_{2,3}\equiv\alpha_{2,3}^{(2)}\left(\frac{m_{2,3}\,T_d}{2\pi}\right)^{3/2}e^{-\epsilon_{2,3}/T_d}.
\end{equation}
Here, $T_{d}$ stands for the temperature of the dark photon bath, and $\epsilon_{2,3}$ stands for the binding energy of $H_{12}$ and $H_{13}$, respectively, given by $\mu\,\alpha_{d}^{2}/2$, with $\mu$ the reduced mass of the bound state in question. $\beta_{2,3}$ is the recombination rate, which relates to the ionization rate of the excited ($n=2$) state 
\begin{equation}
\alpha_{2,3}^{(2)}=9.78\frac{\alpha_d^2}{m_{\chi_{2,3}}^2}\left(\frac{\epsilon_{2,3}}{T_d}\right)^{1/2}\ln\left(\frac{\epsilon_{2,3}}{T_d}\right).
\end{equation}
The factor $C_{2,3}$ takes into account Peebles' correction to the process, and its value can be approximated as
\begin{equation}
C_{2,3}=\frac{\Lambda_{\alpha(2,3)}+\Lambda_{2\gamma_D(2,3)}}{\Lambda_{\alpha(2,3)}+\Lambda_{2\gamma_D(2,3)}+\beta^{(2)}_{2,3}},\quad\Lambda_{\alpha(2,3)}=\frac{H(3\,\epsilon_{2,3})^3}{(8\pi)^2 n_{2,3}(1-X_{2,3})},
\end{equation}
with $H$ being the Hubble rate. $\Lambda_{2\gamma_D(2,3)}$ stands for the two photon decay rates of $H_{12}$ and $H_{13}$. We estimate the two photon decay rate in our model by rescaling the SM rate $\Lambda_{2\gamma}=8.227$ sec$^{-1}$. As given in equation 2 of~\cite{1951ApJ...114..407S}, the parametric dependence of this rate is given by the ground state energy times $\alpha^{6}$, and we substitute the corresponding values of these quantities in our model to obtain $\Lambda_{2\gamma_D(2,3)}$. The Lyman alpha production rate is $\beta^{(2)}_{2,3}=\beta_{2,3}\,e^{3\epsilon_{2,3}/4T_d}$. When calculating the ionization fractions, we calculate the Hubble expansion rate, keeping in mind that the dark and visible sectors have different temperatures, using the benchmark value $\Delta N_{eff}=0.60$ as discussed above.

In Fig.~\ref{fig:recomb}, we plot $X_{3}$ during $H_{13}$ recombination, for a few representative values of $m_{\chi_{3}}$ and $\alpha_{d}$, for $n_3/n_1=10\%$. The plots on the left and right correspond to the case where $\chi_{1}$ is the heaviest flavor with $m_{\chi_{1}}=25$~GeV, and the case where it is the second heaviest flavor with $m_{\chi_{1}}=1$~GeV, respectively. The relic ionization fraction becomes larger either for a smaller value of $\alpha_d$ or a larger value of $m_{\chi_{3}}$. The residual ionization fraction can be approximated as~\cite{CyrRacine:2012fz}
\begin{equation}\label{eq:approxX3}
X_3\sim 3\times 10^{-6}\left(\frac{T_{d}}{T_{\gamma}}\right)\left(\frac{\alpha_d}{0.02}\right)^{-6}\left(\frac{n_1}{n_3}\right)\left(\frac{m_{H_{13}}}{{\rm GeV}}\right)\left(\frac{\epsilon_{3}}{\rm keV}\right).
\end{equation}
Dark baryon acoustic oscillations end after $H_{13}$ recombination, and the recombination time scale determines the suppression of the power spectrum. Since the $H_{12}$ recombination happens at a much earlier time, it does not have a strong effect on the power spectrum.

\subsection{Dark acoustic oscillations and large scale structure}
We now turn our attention to the evolution equations for the DM and dark radiation (DR) density perturbations in our model, after the $H_{12}$ recombination, but before the $H_{13}$ recombination. Thus the relevant matter degrees of freedom are $\chi_{1}$ (with $X_{1}\sim n_{3}/n_{1}$ during this epoch), $\chi_{3}$ (with $X_{3}\sim 1$) and the bound state $H_{12}$ (with $n_{H_{12}}\sim n_{2}$). As discussed earlier in this section, we treat all the dark matter particles to behave as a single fluid during the dark acoustic oscillations and refer to these degrees of freedom collectively as ``Acoustic DM" (AcDM) that undergo acoustic oscillations. We work in the conformal 
Newtonian gauge \cite{Ma:1995ey}
 \begin{equation}
d s^2 
= a^2(\tau) \, 
  \bigl[ -(1 + 2 \psi) d\tau^2 + (1 - 2\phi) \delta_{ij} d x^i d x^j \bigr] \,,
 \end{equation}
where the fields $\psi$ and $\phi$ describe scalar perturbations on the 
background metric. To linear order in the perturbations, we have
\begin{eqnarray}
\dot{\delta}_{{\rm AcDM}}&=&-\theta_{{\rm AcDM}}+3\dot{\phi} \ ,\label{eq:evolution3}
\\
\dot{\theta}_{{\rm AcDM}}&=&-\frac{\dot{a}}{a}\theta_{{\rm AcDM}}
+\frac{4\rho_{\gamma_D}}{3\rho_{{\rm AcDM}}}n^{f}_{3}(a)\sigma_T\,a(\theta_{\gamma_D}-\theta_{{\rm AcDM}})+k^2\psi,\label{eq:beq}
\label{eq:evolution4}
 \end{eqnarray}
where the derivatives are with respect to $\eta$, the conformal time. $\delta\equiv\delta\rho/\bar{\rho}$ is the perturbation of the energy density, $k$ is the wave number, and $\theta\equiv\partial_iv^i$ is the divergence of the comoving 3-velocity. Since the AcDM components are all non-relativistic at this time, one can ignore the sound speed\footnote{We have verified numerically that this gives a good approximation.}. As long as the momentum transfer rate from the dark Thomson scattering $\gamma_D\chi_3\to \gamma_D\chi_3$ is comparable to Hubble, the density perturbations oscillate with the dark photon perturbation, and structures cannot grow. The cross section is given by $\sigma_{T}=8\pi\alpha_d^2/3m_{\chi_{3}}^2$, and $n^{f}_{3}$ depends on the ionization fraction $X_3$ obtained by solving Eq.~(\ref{eq:ionization}). The dark photon perturbations, including higher modes in the Legendre polynomials, $F_{\gamma_D\ell}$, evolve as~\cite{Ma:1995ey}
 \begin{eqnarray}
\dot{\delta}_{\gamma_D}&=&-\frac{4}{3}\theta_{\gamma_D}+4\dot{\phi},\label{eq:evolution5}
\\
\dot{\theta}_{\gamma_D}&=&k^2\left(\frac{1}{4}\delta_{\gamma_D}-\frac{1}{2}F_{\gamma_D2}\right)+a\,n^{f}_{3}\,\sigma_T(\theta_{{\rm AcDM}}-\theta_{\gamma_D})+k^2\psi,\label{eq:evolution6}
\\
\dot{F}_{\gamma_D2}&=&\frac{8}{15}\theta_{\gamma_D}-\frac{3}{5}k F_{\gamma_D3}-\frac{9}{10}a\,n^{f}_3\,\sigma_TF_{\gamma_D2},\label{eq:evolution7}
\\
\dot{F}_{\gamma_Dl}&=&\frac{k}{2l+1}\left[lF_{\gamma_D(l-1)}-(l+1)F_{\gamma_D(l+1)}\right]-a\,n^{f}_3\,\sigma_TF_{\gamma_Dl},\quad l\geq3\label{eq:evolution8}
\\
\dot{F}_{\gamma_Dl_{\text{max}}}&=&kF_{\gamma_D(l_{\text{max}}-1)}-\frac{l_{\text{max}}+1}{\tau}F_{\gamma_Dl_{\text{max}}}-a\,n^{f}_3\,\sigma_TF_{\gamma_Dl_{\text{max}}}.\label{eq:evolution9}
\end{eqnarray}
Here the $F_{\gamma_Dl}$ are related to the spatial variations in the density fluctuations in the dark photons, in particular $\delta_{\gamma_D} \equiv F_{\gamma_D0}$, $\theta_{\gamma_D} \equiv \frac{3}{4}kF_{\gamma_D1}/4$, and $\sigma \equiv \frac{1}{2}F_{\gamma_D2}$ where $\sigma$ is the shear stress. We truncate the Boltzmann hierarchy at order $l_{\text{max}}=4$, making use of the approximation outlined in Ref.~\cite{Ma:1995ey}\footnote{We have also reproduced the analysis with $l_{\text{max}}=5$ to confirm that our results are not sensitive to the choice of $l_{\text{max}}$.}. The equations are similar to those for the SM photon and baryons.

In the calculation we take $\psi=\phi$ and ignore the correction from free streaming radiation. This approximation is good since our $\Delta N_{eff}$ is much smaller than the number of light neutrinos. Gravity perturbations are sourced by the density fluctuations as described by the Einstein equation,
 \begin{equation}
k^2\psi+3\frac{\dot{a}}{a}\left(\dot{\psi}+\frac{\dot{a}}{a}\psi\right)=-\frac{a^2}{2M_{pl}^2}\sum\rho_{i}\,\delta_{i}, \label{eq:evolution10}
 \end{equation}
where the sum is over the SM photon, the dark photon and the AcDM components. For the initial conditions, the modes that enter the horizon before 
matter-radiation equality satisfy
 \begin{equation}
\delta_{\gamma_D}=\frac{4}{3}\delta_{{\rm AcDM}}=-2\psi,\quad\theta_{\gamma,\gamma_D,{\rm AcDM}}=\frac{k^2\eta}{2}\psi,
 \end{equation}
 and the modes that enter during the era of matter domination satisfy
 \begin{equation}
\frac{3}{4}\delta_{\gamma_D}=\delta_{{\rm AcDM}}=-2\psi,\quad\theta_{\gamma,\gamma_D,{\rm AcDM}}=\frac{k^2\eta}{3}\psi.
 \end{equation}
 We set the initial values of the higher modes 
$F_{\gamma_D\ell\,\geq2}=0$, since these higher angular modes 
quickly damp away when the ${\rm AcDM}$-$\gamma_D$ scattering is efficient. We neglect the 
tilt in the primordial spectrum ($n_s=1$) and take a $k$-independent 
value of $\psi= 10^{-4}$. The final results are independent of the 
precise value of $\psi$ since we are interested in the ratio of the 
matter power spectra with and without the dark acoustic oscillations. In the 
numerical study, we choose the values $h=0.67$, $\Omega_{\gamma} h^2 = 
2.47 \times 10^{-5}$, $\Omega_{\Lambda} h^2 = 0.69$, $\Omega_{b} h^2 = 
2.2\times 10^{-2}$, and $\Omega_{\nu}=0.69 \Omega_{\gamma}$~\cite{Aghanim:2018eyx}.

\begin{figure}
\begin{center}
\includegraphics[width=10cm]{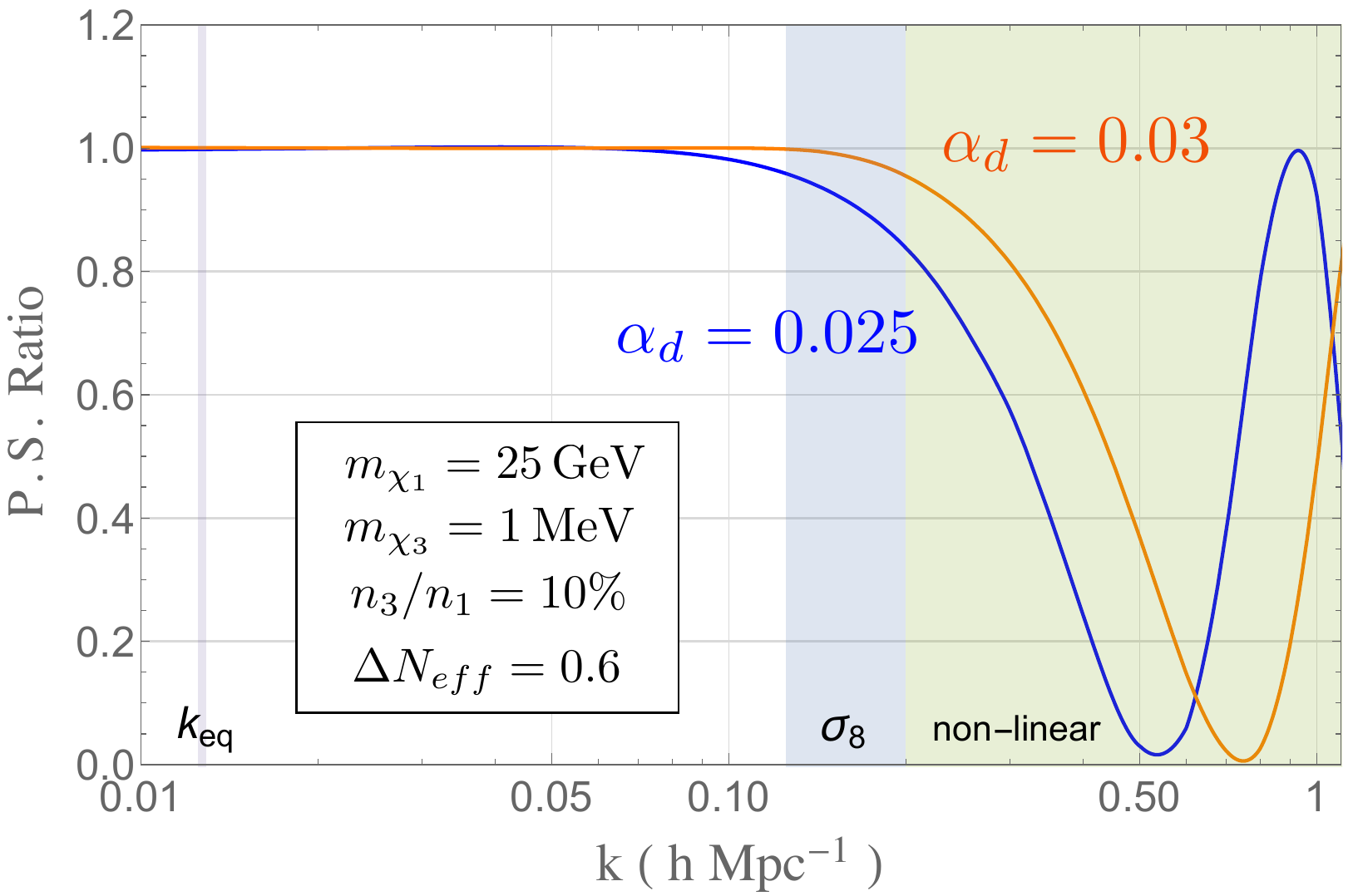}
\caption{Ratio of the matter power spectrum between scenarios with and without dark acoustic oscillations, as is defined in Eq.~(\ref{eq:PSratio}). We use two representative values of $\alpha_{d}$ in the region of interest. The result is obtained by solving the linear order equations listed in the text. The ratio at large $k$ is expected to receive further corrections from non-linear effects.}\label{fig:PSratio}
\end{center}
\end{figure}

After solving this set of differential equations, in order to quantify the importance of dark acoustic oscillations, we compare the DM power spectrum of AcDM to that of collisionless DM with an added non-interacting dark photon component, such that the energy density of dark radiation is equal in both scenarios, and further complications in the fitting of cosmological parameters are avoided:
 \begin{equation}\label{eq:PSratio}
 \frac{P(k)_{\rm AcDM}}{P(k)_{\Lambda{\rm CDM+DR}}}\approx\left(\frac{\delta_{\chi}(k)_{{\rm AcDM}}}{\delta_{\chi}(k)_{\Lambda{\rm CDM+DR}}}\right)^2,
 \end{equation}
where the terms on the right hand side refer only to the nonrelativistic DM component. We show the power spectrum ratio in Fig.~\ref{fig:PSratio} for two representative values of $\alpha_{d}$ in the region of interest. When the $H_{13}$ recombination takes place, the momentum transfer term in Eq.~(\ref{eq:beq}) quickly drops blow the Hubble expansion rate. The density perturbations entering the horizon after this point evolve the same way as they would in the $\Lambda$CDM scenario, and thus the ratio for small $k$ modes asymptotes to $1$. The matter power spectrum receives a suppression for modes that enter the horizon before recombination, thus for a lower $H_{13}$ binding energy (blue curve) there is a larger suppression. This helps to explain the results of low red-shift measurements for $\sigma_{8}$. We estimate the viable parameter region by requiring a $5-15\%$ suppression of the power spectrum at $k= 0.2\,h$ Mpc$^{-1}$ (blue band in figure~\ref{fig:PSratio} and yellow band in figure~\ref{fig:PSscan}). Although in this work we only focus on the suppression of the matter power spectrum in the linear regime, the suppression continues past $k=0.2h$ Mpc$^{-1}$. One thing to notice is that unlike the warm dark matter scenario that can totally eliminate the matter power spectrum at small scales, the suppression due to the dark acoustic oscillations itself oscillates before entering the non-linear regime. As is shown, e.g., in Ref.~\cite{Buckley:2014hja,Bose:2018juc}, the gravitational collapse after redshift $z\lsim 10$ is likely to destroy this oscillation pattern and therefore it reduces the suppression of the power spectrum in the non-linear region. Once the non-linear corrections to the density perturbation are included, the galaxy survey data and Lyman-$\alpha$ observations, which probe the matter power spectrum at even larger $k$-modes, can be used to further constrain dark acoustic oscillations~\cite{Cyr-Racine:2013fsa,Krall:2017xcw, Garny:2018byk,Rivero:2018bcd}, and thereby the parameter space of our model.

\subsection{Scattering between bound states and small scale structure}

After halo formation, the scattering cross section between the non-relativistic $H_{12}$ bound states in the halo is roughly geometric in size, $\sigma\sim (\mu_{12}\alpha_d)^{-1}$. For the region of parameter space we are interested in, where the heaviest DM flavor has a mass of ${\mathcal O}(10)$~GeV and $\alpha_d\sim 10^{-2}$, the resulting cross section over mass ratio $\sigma/m_{H_{12}}\sim 0.1$ cm$^2/$g can be large enough to thermalize the bound states and change the DM density in the inner part of the halo. It was pointed out in Ref.~\cite{Boddy:2016bbu} that dark hydrogen with a similar range of mass and couplings can explain the low DM density cores observed in small galaxies. Moreover, if the hyperfine splitting of the bound state
\begin{equation}
E_{hf}=\frac{8}{3}\,\alpha_d^2\,f(R)^{-1}E_0,\quad R\equiv\frac{m_{\chi_{1}}}{m_{\chi_{2}}},\quad f(R)=R+2+R^{-1}-\frac{\alpha_d^2}{2}
\end{equation}
normalized by the energy scale $E_0=\alpha_d^{2}\;\mu_{12}$ is comparable to the kinetic energy of the bound state, the effect of the inelastic hyperfine upscattering is to enhance the cross section of $H_{12}$ self-interactions. The velocity dependence in this process makes the self-interaction cross sections in dwarf halos larger than that in cluster halos, giving the correct $\sigma/m$ ratio to solve the mass deficit problem in galaxy clusters. 

In order to show that the same mechanism also works in the region of parameter space that is of interest to us, we adopt the best fit value from ref.~\cite{Boddy:2016bbu}
\begin{equation}\label{eq:hfenergy}
E_{hf}=10^{-4}E_0
\end{equation}
and we limit ourselves to the range of $\alpha_d$ in Fig.~3 of~\cite{Boddy:2016bbu}, where both the dwarf and cluster data can be explained. Since we want the $H_{12}$ self-interaction to solve the small scale structure problem, we choose $n_{3}/n_{1}=0.1$ as our benchmark value such that $H_{12}$ and not $H_{13}$ is the dominant component of DM, while the $n_3$ number density is not unnaturally small (see figure~\ref{fig:xrpdf}), and also not too small to maintain the thermal equilibrium in the early universe that is responsible for suppressing structure formation and addressing the $\sigma_{8}$ discrepancy. 

Even though it constitutes a smaller fraction of the DM energy density, we need to assess whether the $H_{13}$ bound state may still play a role in halo formation. In particular, the $H_{13}$ bound state has a much larger radius than $H_{12}$ due to the smallness of $m_{\chi_{3}}$, and therefore scattering with $H_{13}$ could potentially change the desired core size of the $H_{12}$ halo. However, we find that this is not the case. For the parameter range we consider, the inverse Bohr radius of $H_{13}$ ($\sim10$ keV) is still too small compared to to the value that would result in sufficient momentum transfer for keeping $H_{12}$ atoms in thermal equilibrium both at dwarf galaxies ($\sim$ MeV) and galaxy clusters ($\sim 100$ MeV). Thus the scattering process most efficient for momentum transfer is off the $\chi_{1}$ particle (i.e. the ``nucleus'') inside $H_{13}$, with a cross section comparable to the $H_{12}$ self-scattering. Therefore the geometric size of $H_{13}$ does not lead to an enhancement, and the $H_{12}$ isothermal profile is not significantly affected.

As structures form, the bound states fall into the overdense region and their gravitational potential energy is converted into kinetic energy, resulting in shock-heating to a temperature~\cite{Fan:2013yva}
\begin{equation}\label{eq:Tgal}
T_{gal}\approx  0.86\,{\rm keV}\frac{\mu}{10\,{\rm GeV}}
\end{equation}
for a Milky Way sized galaxy with halo mass $10^{12}M_{\odot}$ and radius $110$ kpc. Here $\mu$ is the total mass of all degrees of freedom that contribute to $\rho_{DM}$ divided by their total number density, given by
\begin{equation}
\mu=\frac{(n_{2}-n^{f}_{2}) m_{H_{12}} + (n_{3}-n^{f}_{3})m_{H_{13}} +\sum_{i=1}^{3} n^{f}_{i}m_{\chi_{i}}}{(n_{2}-n^{f}_{2})+(n_{3}-n^{f}_{3})+\sum_{i=1}^{3} n^{f}_{i}}.
\end{equation}
If $T_{gal}$ is higher than the binding energy, bound states can dissociate. While the more tightly bound $H_{12}$ does not dissociate in the region of parameter space that is of interest to us, if $H_{13}$ ``reionizes'' in this fashion, the scattering process $\chi_3 \chi_1 \to\chi_3 \chi_1 \gamma_D$ can lead to efficient cooling through bremsstrahlung. For simplicity, when checking for the ionization of $H_{13}$, we take the initial condition to be $n^{f}_{i}=0$. In the parameter region where $H_{13}$ reionizes, we then recalculate $T_{gal}$ with $n^{f}_{1}=n^{f}_{3}=n_{3}$ (fully reionized $H_{13}$) to use in the estimate for the cooling process. An estimation of the cooling time scale through this process is given by~\cite{Fan:2013yva}
\begin{equation}\label{eq:tcool}
t_{brem}\sim 6\,{\rm Gyr}\left(\frac{0.02}{\alpha_d}\right)^3\left(\frac{0.1}{n_3/n_1}\right)\left(\frac{\mu}{10\,{\rm GeV}}\right)^{\frac{1}{2}}\left(\frac{m_{H_{12}}}{10\,{\rm GeV}}\right)\left(\frac{m_{\chi_{3}}}{1\,{\rm MeV}}\right)^{\frac{3}{2}}.
\end{equation}
The emitted dark photon can have a free streaming length much larger than the size of the halo, leading to halo cooling. If $t_{brem}$ is much shorter than the age of the Milky Way ($T_{MW}$), a dark disc may form. Recently, results from the GAIA survey~\cite{2016A&A...595A...1G} have been used to set an upper bound on the fraction of the DM that can be contained in a dark disc at $\sim 1\%$~\cite{Schutz:2017tfp,Buch:2018qdr}. A detailed study of the cooling process and the merger history of sub-halos is beyond the scope of this paper, therefore as we scan through the parameter space of our model, we will use $t_{brem} / T_{MW}$ as a conservative indicator of whether there is a significant probability of dark disc formation. In Fig.~\ref{fig:PSscan}, the region where $H_{13}$ can be reionized due to shock heating is shown below the red-dotted curve, and the region where the condition for efficient bremsstrahlung cooling is satisfied is shown above the red dashed curve, resulting in the red shaded region where both conditions are satisfied and where a dark disc may form.

\subsection{Reconciling the large and small scale structure problems}
\begin{figure}
\begin{center}
\includegraphics[width=7.4cm]{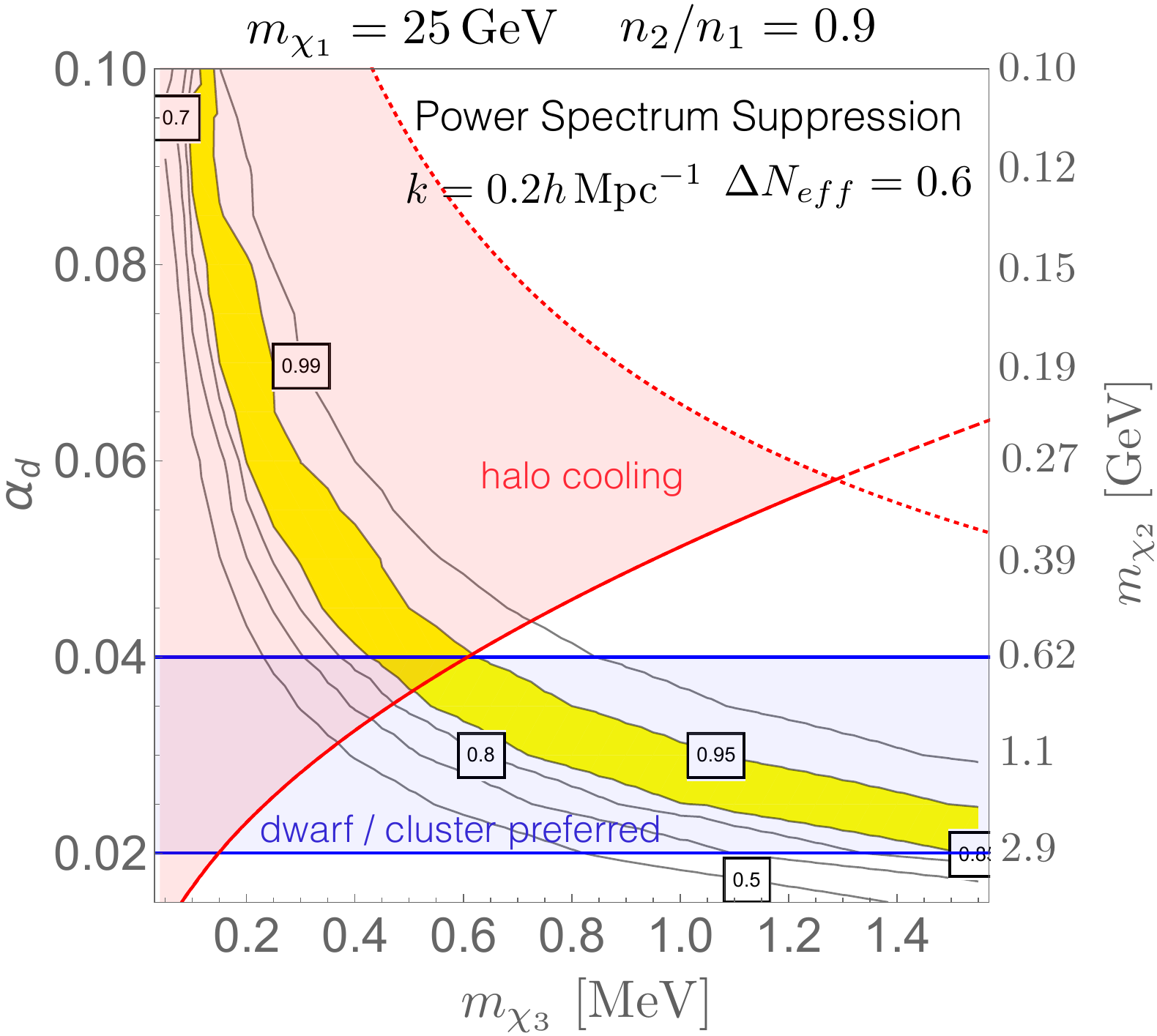}\,\,\,\,\,\,\includegraphics[width=7.4cm]{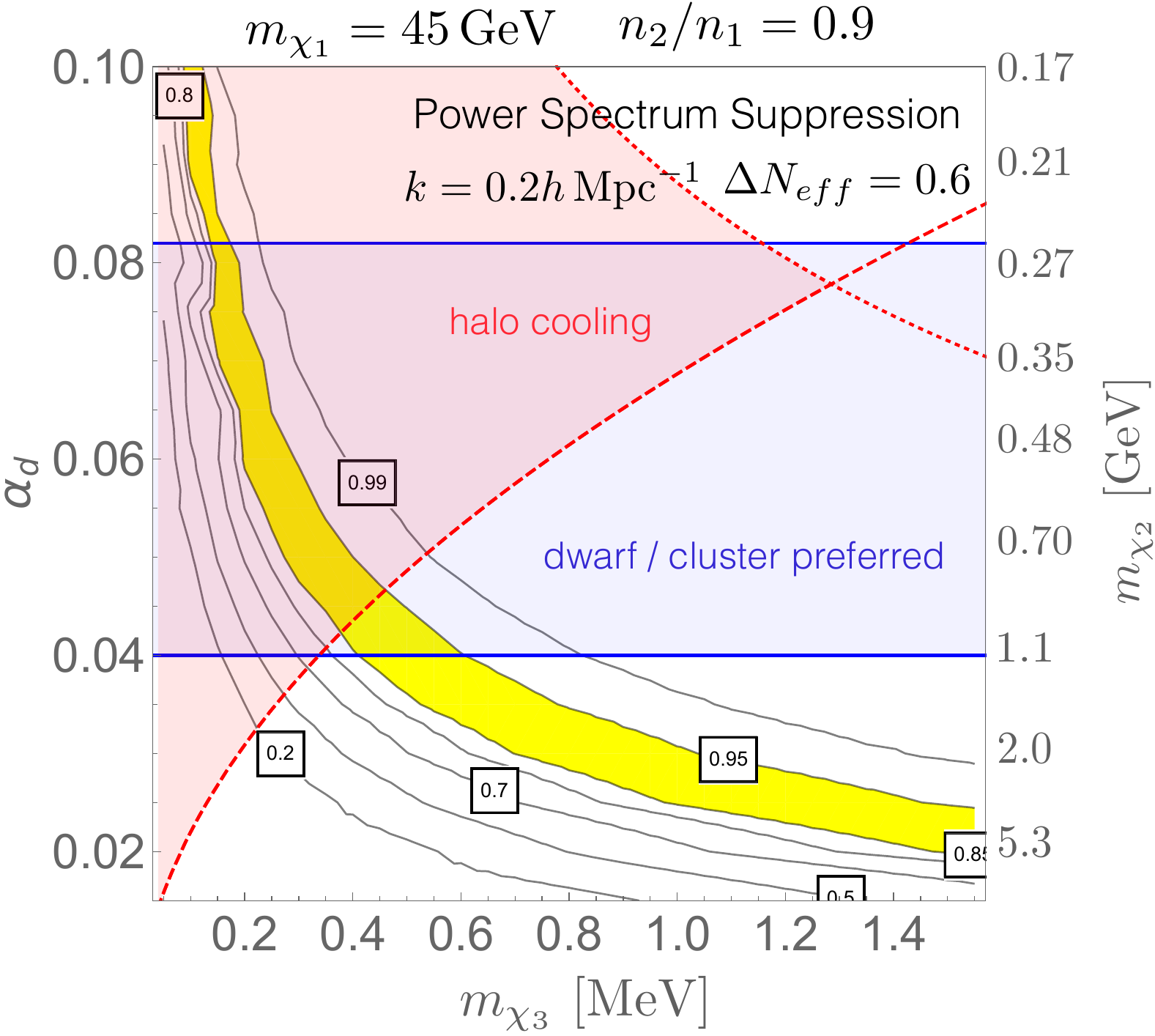}
\\\vspace{1em}
\includegraphics[width=7.4cm]{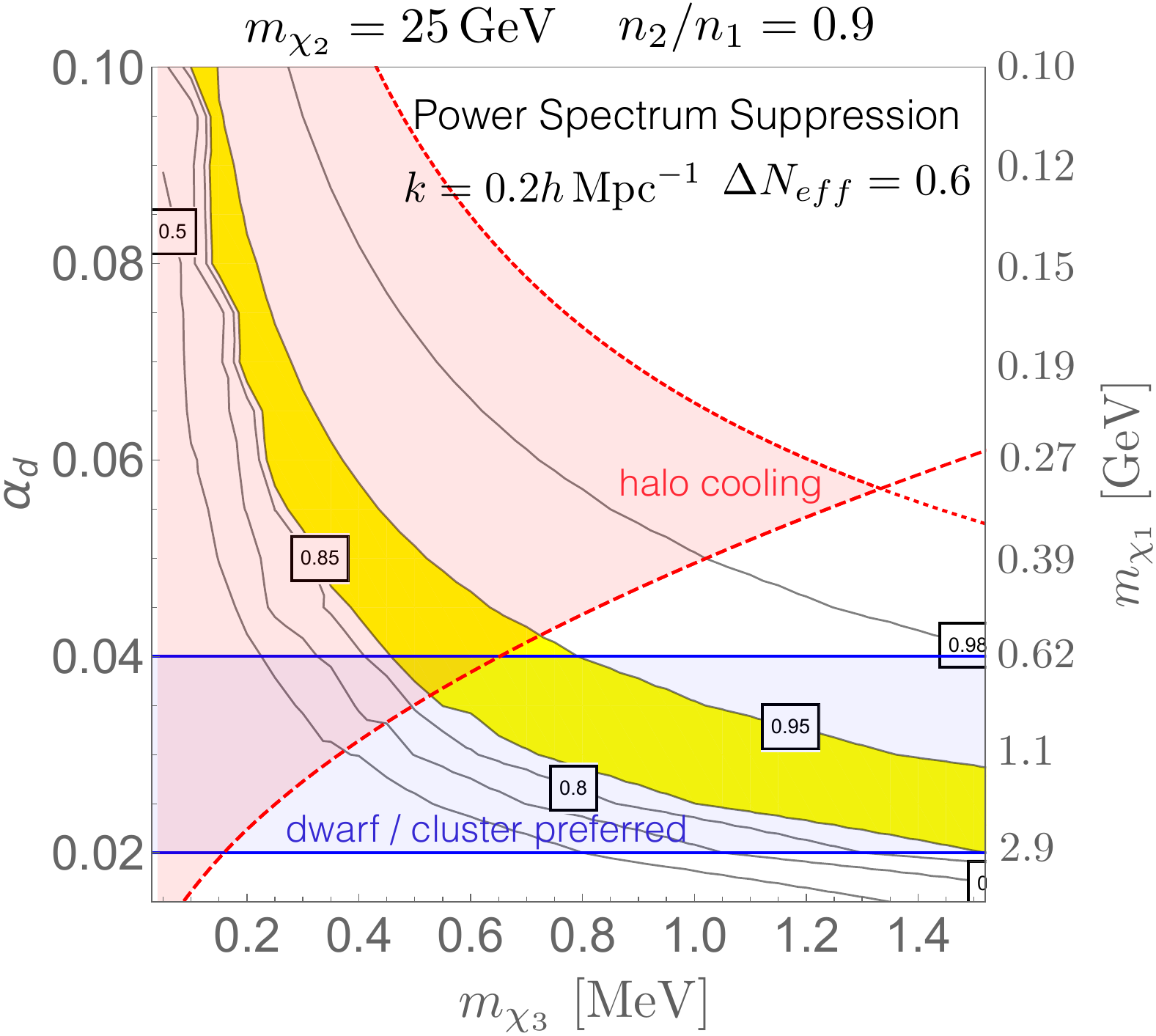}\,\,\,\,\,\,\includegraphics[width=7.4cm]{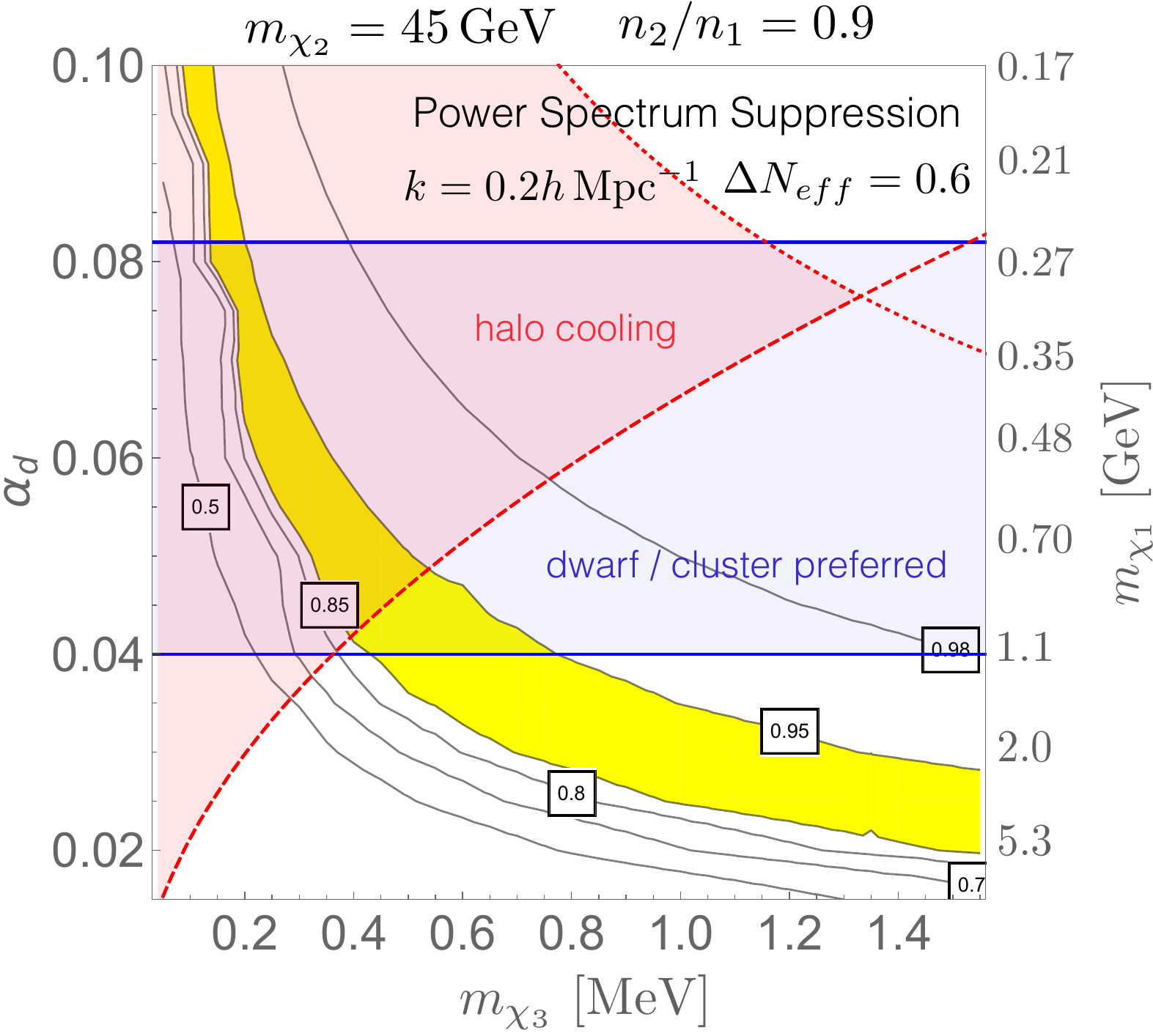}
\caption{The combination of our results, for a few representative parameter choices. Plots on the left (right) column have the mass of the heaviest DM flavor at 25 (45)~GeV. $n_2/n_1$ is chosen to be 0.9 in all plots. In the upper row, $\chi_{1}$ is the heaviest flavor, so its mass is fixed, while the mass of $\chi_{2}$ varies according to the $E_{hf}/E_0=10^{-4}$ benchmark we have adopted (as indicated by the frame labels on the right-hand side of the plots). In the lower row, the roles of $\chi_{1}$ and $\chi_{2}$ are reversed. As a function of $m_{\chi_{3}}$ and $\alpha_{d}$, between the yellow shaded contours the matter power is suppressed by $5$-$15\%$, which may explain the smaller $\sigma_8$ value from late-time measurements compared to the value obtained by Planck. The blue-shaded bands show the preferred $\alpha_d$ interval where the mass deficit problem from dwarf galaxies to clusters is addressed. The red shaded region is disfavored by dark disc constraints as it allows for efficient cooling of the DM during halo formation. The preferred parameter space is therefore given by the overlap of the yellow and blue bands that is outside of the red shaded region.}\label{fig:PSscan}
\end{center}
\end{figure}

In figure~\ref{fig:PSscan}, we show our combined results for several representative parameter choices, and as a function of $m_{\chi_{3}}$ and $\alpha_{d}$. As mentioned in the previous section, we fix $n_{2}/n_{1}=0.9$. We consider 25~GeV and 45~GeV as the mass of the heaviest DM flavor, considering both possible hierarchies, $m_{\chi_{1}}>m_{\chi_{2}}$ and $m_{\chi_{2}}>m_{\chi_{1}}$.

\begin{itemize}
{\item {\bf Addressing the $\sigma_{8}$ discrepancy:} We calculate the power spectrum ratio of Eq.~(\ref{eq:PSratio}) for $k=0.2h$ Mpc$^{-1}$, which is close to the perturbation mode for the $\sigma_8$ measurement. The contours of the power spectrum suppression depend mainly on the $H_{13}$ recombination time scale, thus a constant power spectrum suppression traces $\alpha_d\propto(m_{\chi_{3}})^{-\frac{1}{2}}$ for a constant $H_{13}$ binding energy, but they do not depend strongly on $m_{\chi_{1}}$. The interesting regions for addressing the $\sigma_8$ problem are shown by the yellow bands. } 
{\item {\bf Small scale structure:} We fix the ratio of the hyperfine splitting to the ground state energy as in Eq.~(\ref{eq:hfenergy}). The mass of the intermediate DM flavor ($\chi_{2}$ for the upper row of plots, and $\chi_{1}$ for the lower row) is then determined at each point. This is indicated by the frame labels on the right-hand side of each plot. The preferred $\alpha_d$ interval for solving both the dwarf and cluster mass deficit problems from ref.~\cite{Boddy:2016bbu} is indicated by the blue shaded bands.}
{\item {\bf Constraints from dark disc formation:} As explained in the previous section, the formation of a dark disc is possible in the red shaded region, which is therefore disfavored.}
\end{itemize}

In summary, the overlap region between the yellow and blue bands that is outside of the red region gives the most preferred parameter space for addressing structure formation puzzles at different scales. This favors the ranges $\alpha_d\approx 0.02$-$0.04$, $20$-$45$ GeV for the mass of the heaviest DM flavor, and the MeV scale for the mass of the lightest flavor. We reiterate that in this study we have taken a relatively simple approach to demonstrate that our model has the potential to solve the relevant structure formation problems; however a more careful study of the cosmological data and the Lyman-$\alpha$ constraints should be performed to fully establish this claim and determine the precise region in the parameter region where all conditions of interest are satisfied.


\section{Conclusions}
\label{sec:conclusions}

We have explored the cosmological and astrophysical implications of a model of Secretly Asymmetric Dark Matter, where flavor-by-flavor asymmetries are generated in the dark sector through the decay of heavy right-handed neutrinos, despite an exact gauged dark $U(1)$. As a result, the total dark charge of the universe is always zero, and the DM flavors have opposite signs of the asymmetry, making it possible for bound states to form. When the heaviest dark matter flavor has a mass of ${\mathcal O}(10)$~GeV, the intermediate flavor has a mass of  ${\mathcal O}(0.1-1)$~GeV, and the lightest flavor has a mass of ${\mathcal O}(1)$~MeV, with an $\alpha_{d}\sim10^{-2}$ and $n_{3} / n_{1} \sim 0.1$, this model can address several outstanding puzzles. In particular, the dark photon as an additional degree of freedom helps resolve the discrepancy between CMB-based and low redshift measurements of $H_{0}$, the resulting dark acoustic oscillations help address the $\sigma_{8}$ problem, and scattering between the bound states $H_{12}$ after halo formation, with an inelastic component through the hyperfine transition, helps resolve issues at the cluster and dwarf galaxy scales. The model is consistent with constraints from short distance physics such as bounds on millicharged DM.

We want to emphasize that while we have chosen the DM to have three flavors for simplicity, the SADM mechanism works for a larger number of flavors as well, resulting in potentially even richer dynamical phenomena due to the increased number of relevant energy scales. We leave the exploration of this possibility to future work. Note also that as experimental sensitivity to millicharged DM increases, especially as the mass threshold of direct detection experiments is lowered, the parameter region of interest to this study should become testable in the not too distant future. Also, if the mass of the mediator $\phi$ is smaller than the benchmark value taken in equation~\ref{eq:decay}, the lifetime of the heavier DM flavors can enter the regime where decaying DM signatures might become observable.

\acknowledgments
The authors express special thanks to Prateek Agrawal and Siva Swaminathan for their contributions at the introductory stages of this study. We also thank Zackaria Chacko and Take Okui for helpful discussions. The research of CK and CT is supported by the National Science Foundation Grant Number PHY-1620610. YT is supported in part by the National Science Foundation Grant Number PHY-1620074 and by the Maryland Center for Fundamental Physics (MCFP). CD is supported in part by the DOE Early Career Grant DE-SC0019225. CK and YT thank the Aspen Center for Physics, supported by National Science Foundation grant PHY-1607611, where part of this work was performed.

\end{document}